\documentclass[9pt,twocolumn,twoside]{osajnl}

\journal{ol} 
\usepackage{xfrac}

\setboolean{shortarticle}{true}

\title{Strong coupling between a plasmonic Fano resonance and anapole states in a metallic-dielectric antenna}

\author[1]{T. C. HUANG}
\author[1]{B. X. WANG}
\author[1,*]{C. Y. ZHAO}

\affil[1]{Institute of Engineering Thermophysics, Shanghai Jiao Tong University, Shanghai 200240, China}

\affil[*]{Corresponding author: changying.zhao@sjtu.edu.cn}

\begin{abstract}
In the quest to enhance light-matter interactions and miniaturize photonics devices, it is crucial to create a strong field enhancement with lower material losses. Here we combine a plasmonic Fano resonance supported by the silver cluster and anapole states realized by the silicon disk to create a larger field enhancement with less loss through a strong coupling effect. Besides, by varying the gap size we find that the resonances wavelength and the Rabi-splitting can be tuned over a wide range of wavelength, which can achieve a giant splitting energy over 300 meV. We further demonstrate that  it is the interference of magnetic currents loops which induces the strong coupling. Due to the strong coupling, the hybrid antenna can provide both larger decay rate and radiative decay rate, which makes it promising for high-performance miniaturized optical devices. 
\end{abstract}

\setboolean{displaycopyright}{true}

\begin{document}

\maketitle

Strong coupling between nanoresonators renders large coherent energy exchange and light-matter enhancements \cite{torma2014strong,villarroya2013sumoylated}, which paves the way for modern nanophotonic devices. Unlike weak coupling that only provides modification in spontaneous decay, strong coupling enable the hybrid of two separation eigenstates due to the energy exchange of two constituents or modes, which exceeds the rate of electromagnetic mode damping \cite{torma2014strong}. Since a strong field confinement is required, plasmonic nanostructures have been regarded as promising platforms for strong coupling. It has been reported that larger quality factor resulted from subradiant modes supported in plasmonic structures can enhance the strength of coupling and enlarge the Rabi-splitting \cite{symonds2008particularities,cade2009strong}.Even though the coupling between pairs of plasmonic resonators can induce a gap plasmon mode  \cite{doi:10.1063/1.2056594,doi:10.1063/1.2149971} and reduce the losses in light propagation, which has been used to guide waves. The effect of non-radiative losses is still significant owing to the extremely strong near-field enhancement and small mode volume induced by localized surface plasmon resonances (LSPRs) in plasmonic structures \cite{sancho2012dark,luk2010fano}, which goes against its applications of plasmonic nanostructures \cite{zeng2014metallo}, such as quantum information operations \cite{yoshie2004vacuum} and fluorescence enhancements \cite{regmi2015nanoscale}. 

As the counterpart of plasmonic nanostructures, dielectric nanostructures with high refractive index have gained much attention recently due to their characteristic of magnetic responses and low instinct losses. Recent years, the destructive interference state of toroidal and electric dipoles, namely an anapole state, has been found as a non-radiative mode with high energy confinement, which can be supported by dielectric particles \cite{zenin2017direct,basharin2017extremely,yang2018anapole}. However, the mode volume of dielectric nanostructures can be much larger than that of plasmonic structures, leading to the smaller maximum of field enhancement. Yet the field energy of anapole states are localized inside the particles \cite{zenin2017direct}, which is not satisfying for enhancing light-matter interactions in practical applications. To take full advantages of both the LSPRs and anapole states, these modes can be designed to provide a larger field enhancement with less non-radiative loss by means of the strong coupling effect. Moreover, the coupling frequency in plasmonic-dielectric structures can be modulated efficiently due to the tunability of both resonances. 
\begin{figure}[htbp]
\centering
\includegraphics[width=\linewidth]{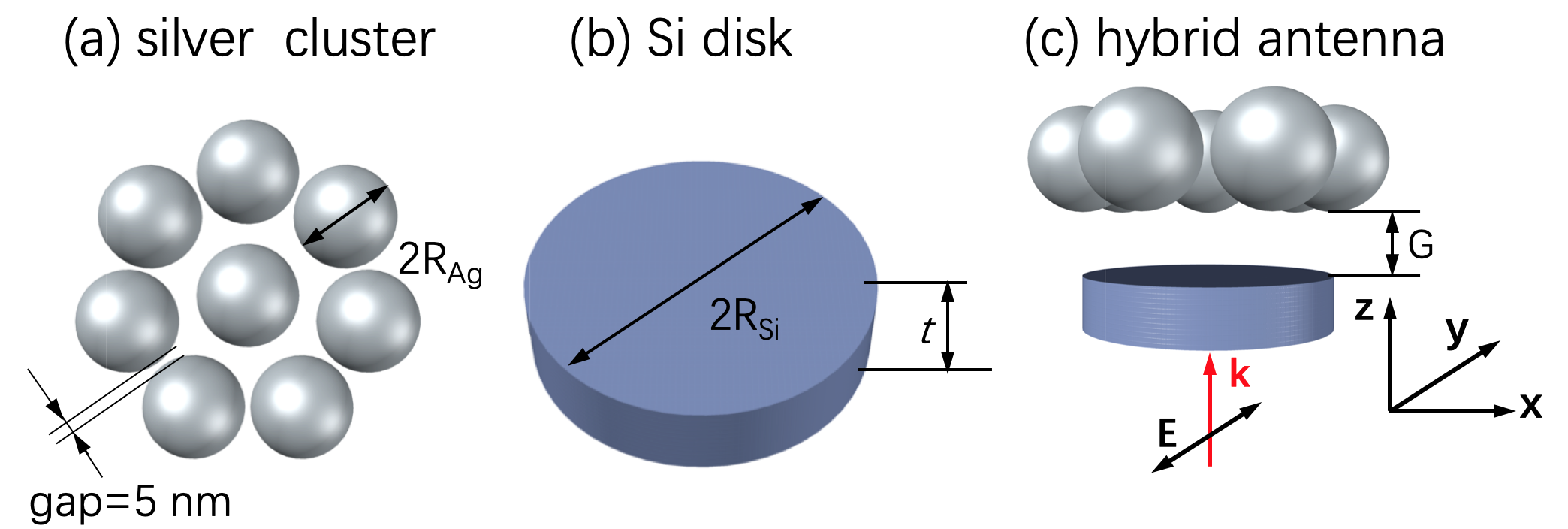}
\caption{Structural view of the proposed (a) silver oligomer cluster, (b) Si disk, and (c) hybrid antenna. ${R}_{\rm{Ag}}$ and $R_{\rm{Si}}$ are the radii of silver particles and Si disk, respectively, $t$ is the thickness of Si disk, and $G$ is the distance between Ag cluster and Si disk. }
\label{fig1}
\end{figure}

In this letter, we propose a hybrid nanoantenna consisting of a plasmonic oligomer cluster and a dielectric nanodisk to realize the strong coupling of two subradiant resonance modes. The coupled modes are found to be effectively tuned in a wide frequency, which can be used to enlarge the Rabi-splitting. We also demonstrate that the underlying mechanism of the coupling between two dark modes is related to the interference of magnetic loops induced in the plasmonic cluster and dielectric disk. Besides, the local density of states and decay rate of this hybrid nanoantenna can be evidently enhanced due to the strong coupling, which makes it potentially applicable for an enhancement of fluorescence \cite{regmi2015nanoscale}, third-harmonic generation \cite{liu2017high,liu2011linear}, surface-enhanced Raman scattering \cite{hwang2018plasmon}, etc.

A silver nanoparticle cluster, as shown in Fig. \ref{fig1} (a), is employed to support a Fano resonance without symmetry breaking \cite{lassiter2012designing}, which origins from the interfering of surface plasmon polaritons in these identical nanoparticles with radius $R_{\rm{Ag}}$ =50 nm and interparticle separation 5 nm. To be specific, this silver oligomer can be regarded as one center particle surrounded by seven ring particles, which support two eigenmodes called center particle mode and ring mode \cite{dregely2011excitation}, respectively. And the coupling of the eigenmodes leads to a dark mode with longer lifetime and a bright mode with larger scattering cross section. The linewidths of bright and dark modes can be obtained by fitting the scattering cross section $\sigma(\omega)=|s(\omega)|^2$ with a two-oscillator Fano interference model \cite{liu2018resonance}, 
\begin{equation}
s(\omega )={ a }_{ r }+\sum _{ m } \frac { { b }_{ m }{ \Gamma  }_{ m }{ e }^{ i{ \phi  }_{ m } } }{ \omega -{ \omega  }_{ m }+i{ \Gamma  }_{ m } } ,
\label{eq1}
\end{equation}
where $a_r$ is the background amplitude, $b_m$, $\Gamma_m$, $\phi_m$, and $\omega_m$ denote, respectively, radiative amplitude, damping, phase, and resonance frequency of the two oscillators characterizing these interfering resonances (m=1, 2). 
\begin{figure}[htbp]
\centering
\includegraphics[width=\linewidth]{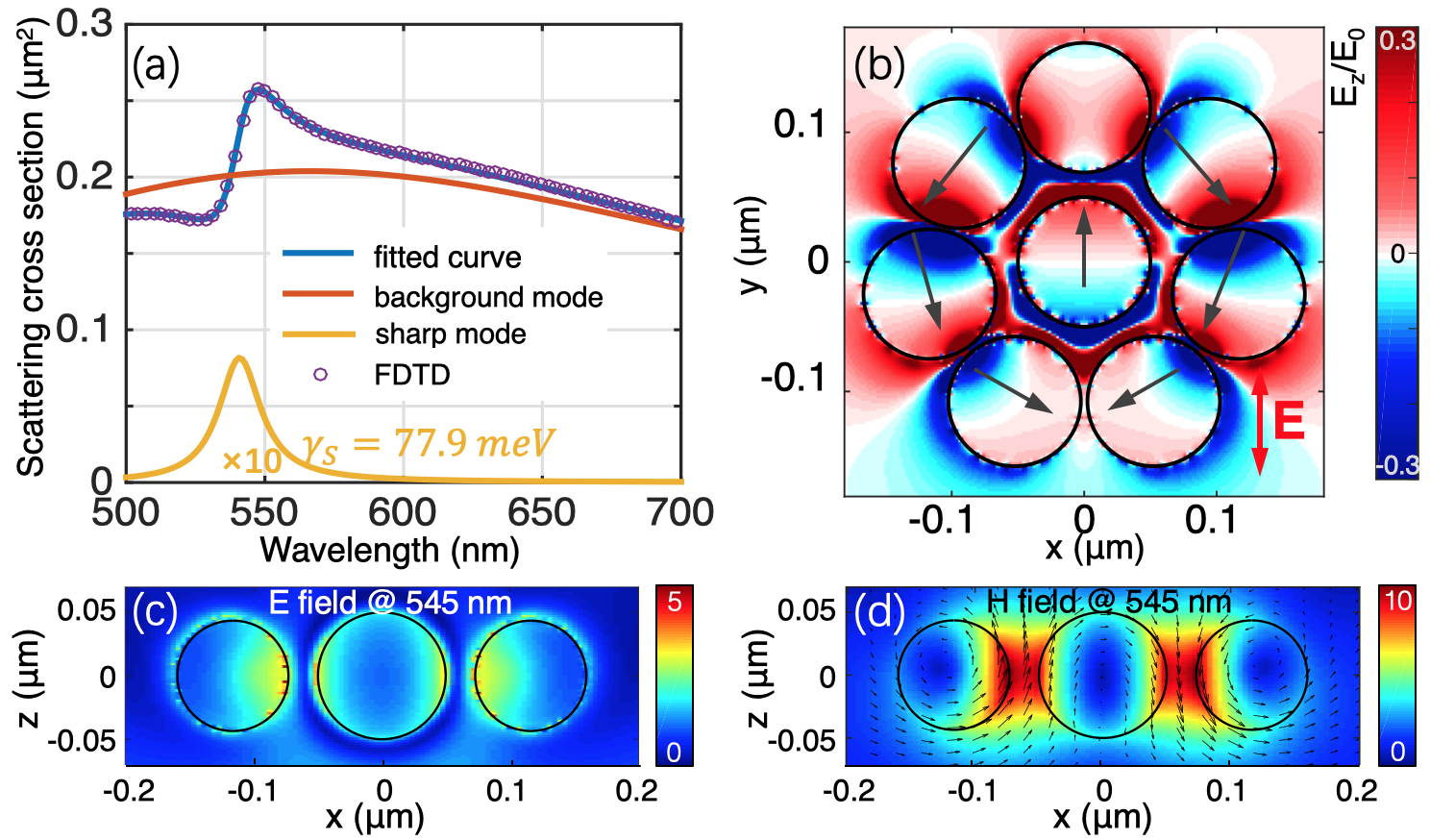}
\caption{(a) The numerically calculated scattering spectra of the proposed silver cluster, and the fitting results with the Fano interference model. The two oscillators used in the Fano interference model are also exhibited. (b)Vertical component of E field when Fano resonance is excited, where polarization in each particle is indicated by the black arrow. (c) and (d) show the electric and magnetic distributions of the subradiant mode (545 nm), respectively. Magnetic field vectors are marked by black arrows.}
\label{fig2}
\end{figure}

Here we exhibit the calculated and fitted scattering spectra of the silver cluster along with the electromagnetic (EM) fields when $\lambda$=545 nm. The dielectric function of silver is adopted from Ref. \cite{johnson1972optical}. As can be seen in Fig. \ref{fig2} (a), the asymmetric fitted curve agrees well with the spectrum calculated by finite-difference time-domain method (FDTD), and the symmetric scattering spectra of two eigenmodes are also exhibited. From the fitting results we know that the linewidth of the dark mode is 77.9 meV, when the linewidth of the bright mode is 1169.2 meV. Based on Fig. \ref{fig2} (b), we verify that the polarization in the oligomer cluster exhibits the characteristic of subradiant mode \cite{lassiter2012designing}, where the center particle mode has opposite polarization to the ring particles mode. The electric and magnetic distributions in $x-z$ plane are displayed in Figs. (c) and (d), respectively. Note that the Fig. \ref{fig2} (d) shows a loop-like magnetic current in the oligomer, whose normal is paralleled with the incident E field. It is worthwhile to mention that the Fano resonances in plasmonic clusters can be effectively tuned by their interparticle separation and particle sizes, which makes them competitive candidates to design a strong coupling.

\begin{figure}[htbp]
\centering
\includegraphics[width=\linewidth]{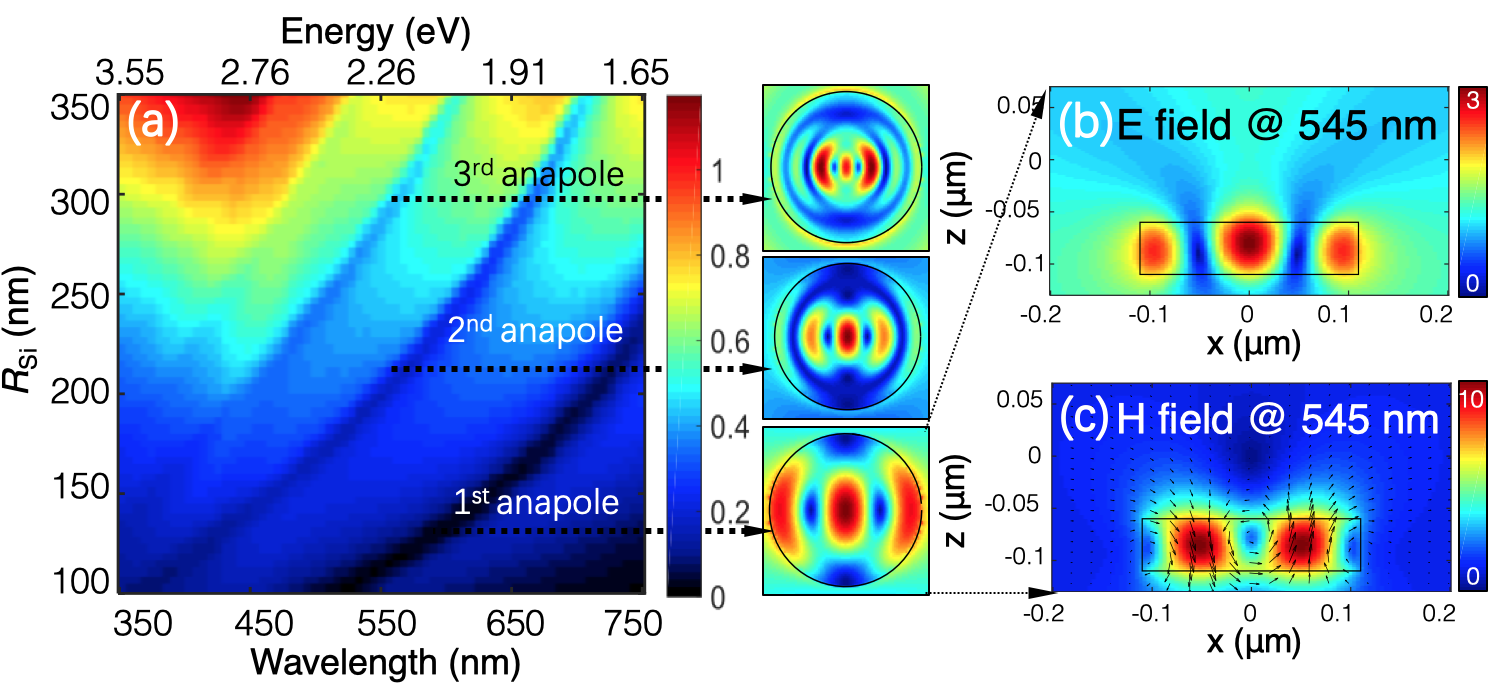}
\caption{(a) Energy diagrams of the Si disk by tuning the radius and the three scattering dips are noted as the first, second and third ordered anapole states. The insets show the electric distribution of these anapole states in $x-y$ plane. (b) and (c) exhibit the $x-z$ plane electric and magnetic distributions of the first ordered anapole state, respectively. The magnetic field vectors are marked by black solid arrows.}
\label{fig3}
\end{figure}

Then we continue to investigate radiationless anapole states excited in silicon disks, which provides a relatively mild field enhancement \cite{yang2018anapole} inside particles (${\left| E/{ E }_{ 0 } \right|}^2 \sim 10$) compared with LSPRs \cite{dregely2011excitation}. Multi-order anapole states can be induced by the destructive interference of toroidal (TD) and electric dipole (ED) modes have been found in previous works \cite{zenin2017direct,yang2018anapole}. To tune the anapole states, we can change the radius of silicon disk $R_{\rm{Si}}$, when the thickness t is fixed at 50 nm. The scattering spectra of silicon disk, whose dielectric constants are adopted from Palik’s book \cite{smith1985handbook}, are shown in Fig. \ref{fig3} (a), and three dark bands can be easily observed. From the electric distributions shown in the insets which are corresponding to the first, second and third ordered anapole states, respectively, we can notice that the internal hot spot becomes more confined with the resonance order. In other words, since a higher-order anapole state offers larger field enhancement than the fundamental one, it is even harder for the hot spot to be accessed by nearby emitters such as molecules and quantum dots. Figs. \ref{fig3} (b) and (c) show the EM field distributions of the first-order anapole state in a disk with $R_{\rm{Si}}$=110 nm, which display the energy enhancement distributions and the TD induced magnetic loop in $x-z$ plane. Note that the magnetic fields vectors of the silicon disk are similar to that of silver cluster (Fig.\ref{fig2} (d)).

\begin{figure}[htbp]
\centering
\includegraphics[width=\linewidth]{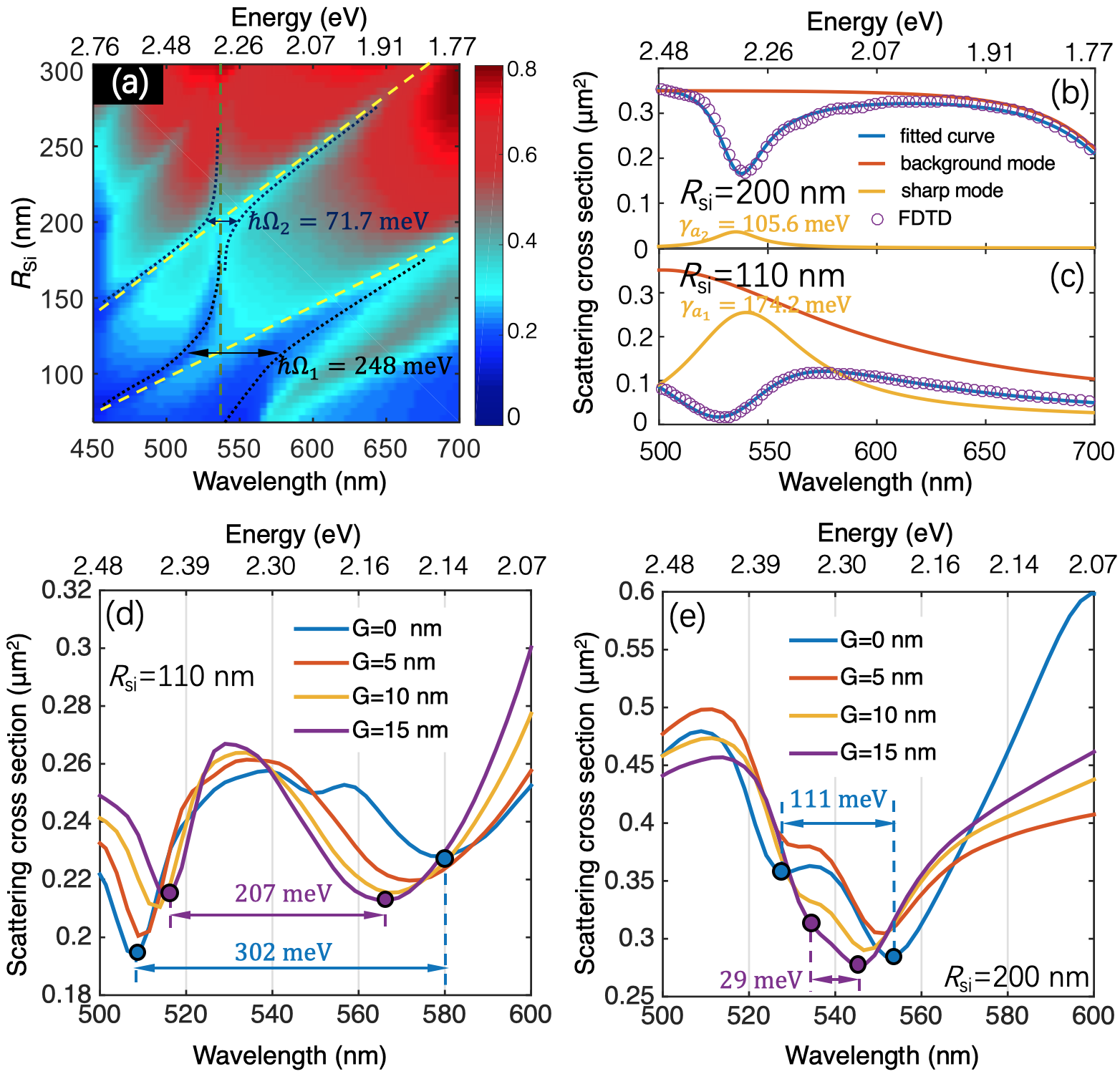}
\caption{(a) Energy diagrams of the disk-cluster hybrid antenna with changing $R_{\rm{Si}}$, where the yellow and green dashed lines denote the evolution of the anapole and the LSPR modes, respectively, and two anti-crossing behaviors can be noticed. (b) and (c) show the fitted oscillating curves of anapole resonances in the Si disk with $R_{\rm{Si}}$=110 nm and 200 nm, respectively. (d) and (e) exhibit the scattering spectra of silicon disk with distance $G$= 0, 5, 10 and 15 nm, when $R_{\rm{Si}}$=110 nm and 200 nm, respectively.}
\label{fig4}
\end{figure}

In order to take advantages of the large field enhancement provided by silver cluster and silicon disk and further make the hot spot more accessible, we combine both structures and produce the energy exchanges between LSPR and anapole states. Here we investigate the scattering spectra of the proposed hybrid structure (shown in Fig. \ref{fig1} (c)) with gap distance $G$=10 nm and $R_{\rm{Si}}$ ranging from 70 nm to 300 nm to determine the coupling strength between these resonance modes. From Fig.\ref{fig4} (a) we can observe two distinguishable anti-crossing behaviors at around 545 nm when  =110 and 200 nm, respectively. From previous researches \cite{torma2014strong,liu2018resonance}, we know that there is another condition must be satisfied to determine the strong coupling, which is that the Rabi-splitting exceeds the rate of electromagnetic mode damping. In other words, the Rabi-splitting energy $\hbar \Omega $ and linewidths of the two coupling modes must satisfy the criteria $\hbar \Omega >({ \gamma  }_{ s }+{ \gamma  }_{ a })/2$,  where ${ \gamma  }_{ s }$ and ${ \gamma  }_{ a }$ are the linewidths of the dominated dark mode in silver cluster and anapole in silicon disk,respectively. 

The Rabi-splitting energy can be determined from Fig. \ref{fig4} (a) by determining the energy shift when $E_S$=$E_a$=2276.5 meV, where $E_S$ and $E_a$ are, respectively, the energies of the uncoupled LSPR and anapole states. Meanwhile, the linewidths of anapole states can be obtained by the coupled harmonic oscillator mode shown in eq. \ref{eq1} \cite{liu2018resonance}, which are exhibited in Fig. \ref{fig4} (b) and (c). Hence, we know that the linewidth of the first and second ordered anapole states are $\gamma_{a1}$=174.2 and $\gamma_{a2}$=105.6 meV, respectively, which also suggests that the q factor increases with the order of anapole states. Since the Rabi-splitting energy are marked in Fig. \ref{fig4} (a) as $\hbar\Omega_1$=248 meV and $\hbar\Omega_2$=71.7 meV, according to the strong coupling criteria, we know that only the first ordered anapole state ($R_{\rm{Si}}$ = 110 nm) and the LSPR mode can create a strong coupling behavior in the hybrid nanoantenna with $G$=10 nm, when the other coupling ($R_{\rm{Si}}$ = 200 nm) is regarded as a weak coupling.

The effect of coupling distance $G$ on the splitting energy is further investigated. Here we calculate the cases with $G$=0,5,10, and 15 nm when $R_{\rm{Si}}$=110 and 200 nm. From Fig.\ref{fig4} we can observe that a smaller $G$ leads to larger Rabi-splitting energy, which can even reach 302 meV when $R_{\rm{Si}}$=110 nm and $G$=0 nm. At the same time, we can realize a strong coupling between the second order anapole mode and LSPR mode since the new splitting energy $\hbar\Omega_2$=111 meV when these two parts approaching each other. Moreover, the coupling distance $G$ can provide an effective method to tune the frequency of coupled modes.

\begin{figure}[htbp]
\centering
\includegraphics[width=\linewidth]{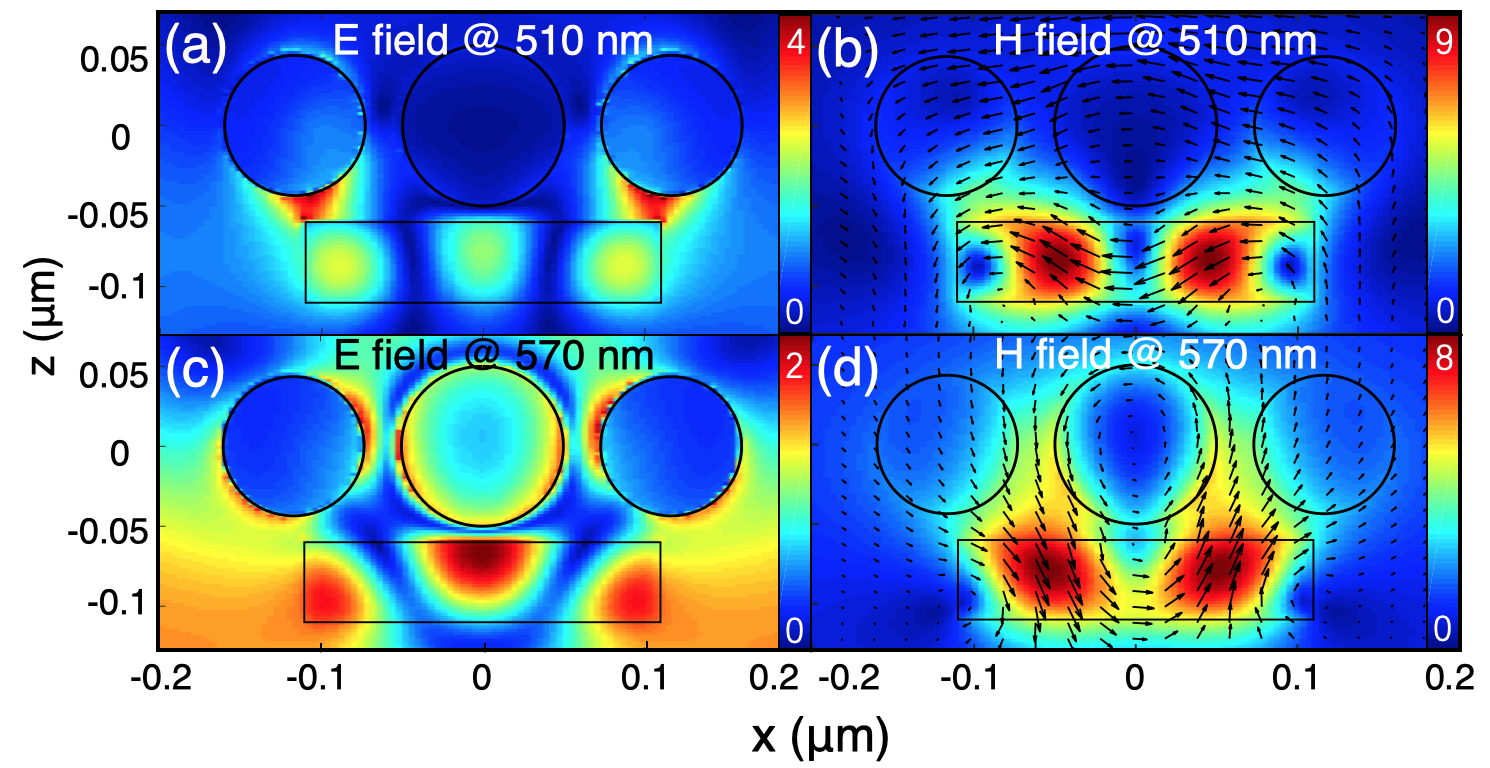}
\caption{The $x-y$ plane EM distributions of strong-coupling-induced (a-b) anti-bounding and (c-d) bounding state in the hybrid nanoantenna with $R_{\rm{Si}}$=110 nm, $G$=10 nm.  The magnetic field vectors are marked by black arrows.}
\label{fig5}
\end{figure}

In order to better understand the nature of strong coupling between the plasmonic Fano resonance and anapole state in hybrid antenna. We present the EM field distributions of the two coupled modes resulted from LSPR and first-order anapole state in Fig. \ref{fig5}. Compared with uncoupled subradiant states in both components, as shown in Figs. \ref{fig2} and \ref{fig3}, it can be easily observed that the electric energy in the upper branch (535 nm) is no longer localized at the inside of silver cluster or silicon disk, but in the gap between them. This phenomenon is analogous to the localized mode in gap plasmon waveguides \cite{doi:10.1063/1.2056594,doi:10.1063/1.2149971}, which reduces the Ohmic loss and makes the hot spot more accessible in practical applications. On the other hand, the EM energy in the lower branch intends to be confined inside the structure instead of vacuum. Besides, the coupling nature in the hybrid nanoantenna can be elucidated by the magnetic field vectors shown in Figs. \ref{fig5} (b) and (d). The anti-bounding pattern of the magnetic currents induced in both structures is exhibited in Fig. \ref{fig5} (b), which shows the coupling of clockwise and anti-clockwise magnetic field vectors. Oppositely, the magnetic field vectors are both anti-clockwise and form a magnetic loop in Fig.\ref{fig5} (d), which shows the characteristic of a bounding state. Similar to Fano resonances in previous researches \cite{luk2010fano,bachelier2008fano,chong2014observation}, the EM fields are more confinement in the anti-bounding state than that in the bounding state. However, the strong coupling in this hybrid antenna results from the coupling of two subradiant modes, which leads to an unusual subradiant bounding state. On the basis of results, the anti-bounding coupled mode in 535 nm is more suitable for achieving stronger field enhancement in practical applications.

To confirm the advantage of the anti-bounding state resulted from strong coupling in near-field enhancement compared with uncoupled situations.We investigate the decay rate of the silver cluster, silicon disk, and their hybrid structure, which is a vital parameter in near-field enhancement. The decay rate is closely related to local density of photonic states \cite{novotny2006principles}, which can be calculated by $\rho_j (r_0,\omega)=(6\omega/{\pi}{c}^2)\cdot \Im \left[    \hat{{G}_{jj}} (r_0,r_0,\omega)\right]$, where $\hat{G}$ is the dyadic Green’s function, $\omega$ is the frequency in vacuum, and $r_0$ is local position, when $j=(x,y,z)$. Note that in this letter we only consider the strong coupling excited by an incident E field which is perpendicular to the $z$ axis. Since the partial LDOS in x and y directions are almost equal from calculation results that are not exhibited in this letter, we can only take partial LDOS in $x$ direction into account. Then the partial decay rate can be obtained by ${ \gamma  }_{ x }=(\omega \pi /3\hbar \epsilon _{ 0 })\cdot { \left| \mu  \right|  }^{ 2 }{ \rho  }_{ x }({ r }_{ 0 },\omega )$, where $\mu$ is the dipole moment, and $\epsilon_0$ is the permittivity in vacuum. For some applications like florescence enhancements, the radiative decay rate is also required to exhibit the effect of near-field quenching.

\begin{figure}[htbp]
\centering
\includegraphics[width=\linewidth]{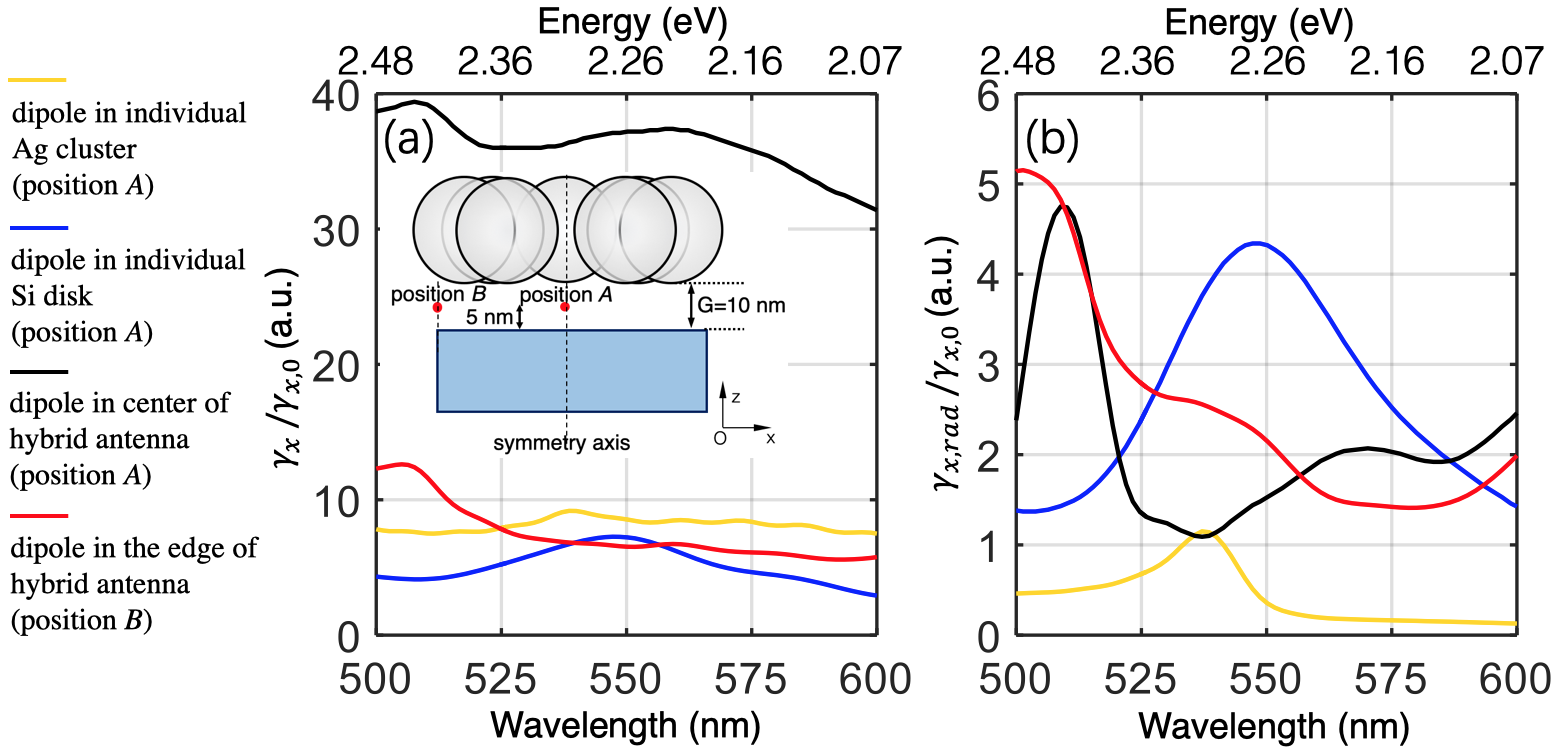}
\caption{(a) and (b) illustrate the enhancement of decay rate and radiative decay rate when the electric dipole with electric fields paralleled with $y$ axis. The inset shows the positions of electric dipole sources, which are marked as red dots.}
\label{fig6}
\end{figure}

To obtain the largest decay rate of hybrid antenna, the dipole emitters are placed at the hot spots, whose location are marked in the inset of Fig. \ref{fig6} (a). The position $A$ locates at the symmetry axis of silver cluster/dielectric disk/hybrid antenna, while the projection of $B$ in $x-y$ plane overlaps with the edge of Si disk. Both $A$ and $B$ are 5 nm away from the nearest surface. From Fig. \ref{fig6} (a) we can observe that the decay rate inside the center hybrid structures ($A$) can be greatly enhanced by more than 4 times than that in isolated silver cluster or Si disk. At the same time, the decay rate at the edge of the gap ($B$) also gets enhanced due to the anti-bounding state. Hence, it is proved that the strong coupling leads to significant enhancement in light-matter interactions, especially the anti-bounding state in 510 nm. Whereas, since the increasing decay rate always accompanied by a greater non-radiative decay rate caused by the losses in materials, which can be further obtained from the far-field power $P_f$ by applying FDTD method. From Fig. \ref{fig6} (b) we know that the loss of metallic structures is much larger than that of dielectric ones, which results in a lower radiative decay rate in isolated silver cluster. However, with the help of strong coupling between LSPRs and anapole states, the hybrid antenna shows larger radiative decay rate than that of an isolated Si disk, which suggests the potential of this hybrid antenna in improving efficiency of applications like enhancing fluorescent signals and quantum emitters. 

In summary, we demonstrate a hybrid metallic-dielectric antenna, which can support the strong coupling between a plasmonic Fano resonance and anapole states. The resonance wavelength and Rabi-splitting energy can be effectively tuned by varying the distance between the silver cluster and the silicon disk. In this way, we can obtain not only a giant Rabi-splitting over 300 meV but also the strong coupling between a second-order anapole state and a plasmonic Fano resonance. From the EM distributions of the coupled modes we find that the modes created by strong coupling is resulted from the anti-bounding and bounding of magnetic loops in both parts in the hybrid antenna. Compared with the bounding state resulted from strong coupling and both original modes, the anti-bounding state induces a stronger electric field enhancement inside the gap region. The subradiant feature of original modes and the EM spatial distributions of the anti-bounding state suppress the non-radiative loss,which results in both the decay rate and radiative decay rate can be efficiently enhanced by the strong coupling. These results suggest that the proposed hybrid nanoantenna with strong coupling effect may pave the way for miniaturized integrated optical devices with high efficiency.\vspace{10pt}\\
{\textsf{\textbf{FUNDING.}}} National Natural Science Foundation of China (51636004, 51476097); Shanghai Key Fundamental Research Grant (18JC1413300, 16JC1403200); Foundation for Innovative Research Groups of the National Science Foundation of China (51521004).

\bigskip

\bibliography{ref}

\begin{thebibliography}{10}
\newcommand{\enquote}[1]{``#1''}

\bibitem{torma2014strong}
P.~T{\"o}rm{\"a} and W.~L. Barnes, {\protect\JournalTitle{Reports on Progress
  in Physics}} \textbf{78}, 013901 (2014).

\bibitem{villarroya2013sumoylated}
C.~Villarroya-Beltri, C.~Guti{\'e}rrez-V{\'a}zquez, F.~S{\'a}nchez-Cabo,
  D.~P{\'e}rez-Hern{\'a}ndez, J.~V{\'a}zquez, N.~Martin-Cofreces, D.~J.
  Martinez-Herrera, A.~Pascual-Montano, M.~Mittelbrunn, and
  F.~S{\'a}nchez-Madrid, {\protect\JournalTitle{Nature communications}}
  \textbf{4}, 2980 (2013).

\bibitem{symonds2008particularities}
C.~Symonds, C.~Bonnand, J.~Plenet, A.~Br{\'e}hier, R.~Parashkov, J.~Lauret,
  E.~Deleporte, and J.~Bellessa, {\protect\JournalTitle{New Journal of
  Physics}} \textbf{10}, 065017 (2008).

\bibitem{cade2009strong}
N.~Cade, T.~Ritman-Meer, and D.~Richards, {\protect\JournalTitle{Physical
  Review B}} \textbf{79}, 241404 (2009).

\bibitem{doi:10.1063/1.2056594}
G.~Veronis and S.~Fan, {\protect\JournalTitle{Applied Physics Letters}}
  \textbf{87}, 131102 (2005).

\bibitem{doi:10.1063/1.2149971}
D.~F.~P. Pile, T.~Ogawa, D.~K. Gramotnev, Y.~Matsuzaki, K.~C. Vernon,
  K.~Yamaguchi, T.~Okamoto, M.~Haraguchi, and M.~Fukui,
  {\protect\JournalTitle{Applied Physics Letters}} \textbf{87}, 261114 (2005).

\bibitem{sancho2012dark}
J.~Sancho-Parramon and S.~Bosch, {\protect\JournalTitle{ACS nano}} \textbf{6},
  8415 (2012).

\bibitem{luk2010fano}
B.~Luk'yanchuk, N.~I. Zheludev, S.~A. Maier, N.~J. Halas, P.~Nordlander,
  H.~Giessen, and C.~T. Chong, {\protect\JournalTitle{Nature materials}}
  \textbf{9}, 707 (2010).

\bibitem{zeng2014metallo}
X.~Zeng, W.~Yu, P.~Yao, Z.~Xi, Y.~Lu, and P.~Wang,
  {\protect\JournalTitle{Optics express}} \textbf{22}, 14517 (2014).

\bibitem{yoshie2004vacuum}
T.~Yoshie, A.~Scherer, J.~Hendrickson, G.~Khitrova, H.~Gibbs, G.~Rupper,
  C.~Ell, O.~Shchekin, and D.~Deppe, {\protect\JournalTitle{Nature}}
  \textbf{432}, 200 (2004).

\bibitem{regmi2015nanoscale}
R.~Regmi, A.~A. Al~Balushi, H.~Rigneault, R.~Gordon, and J.~Wenger,
  {\protect\JournalTitle{Scientific reports}} \textbf{5}, 15852 (2015).

\bibitem{zenin2017direct}
V.~A. Zenin, A.~B. Evlyukhin, S.~M. Novikov, Y.~Yang, R.~Malureanu, A.~V.
  Lavrinenko, B.~N. Chichkov, and S.~I. Bozhevolnyi,
  {\protect\JournalTitle{Nano letters}} \textbf{17}, 7152 (2017).

\bibitem{basharin2017extremely}
A.~A. Basharin, V.~Chuguevsky, N.~Volsky, M.~Kafesaki, and E.~N. Economou,
  {\protect\JournalTitle{Physical Review B}} \textbf{95}, 035104 (2017).

\bibitem{yang2018anapole}
Y.~Yang, V.~A. Zenin, and S.~I. Bozhevolnyi, {\protect\JournalTitle{ACS
  Photonics}} \textbf{5}, 1960 (2018).

\bibitem{liu2017high}
S.-D. Liu, Z.-X. Wang, W.-J. Wang, J.-D. Chen, and Z.-H. Chen,
  {\protect\JournalTitle{Optics express}} \textbf{25}, 22375 (2017).

\bibitem{liu2011linear}
H.~Liu, G.~Li, K.~Li, S.~Chen, S.~Zhu, C.~Chan, and K.~Cheah,
  {\protect\JournalTitle{Physical Review B}} \textbf{84}, 235437 (2011).

\bibitem{hwang2018plasmon}
I.~Hwang, J.~Yu, J.~Lee, J.-H. Choi, D.-G. Choi, S.~Jeon, J.~Lee, and J.-Y.
  Jung, {\protect\JournalTitle{ACS Photonics}} \textbf{5}, 3492 (2018).

\bibitem{lassiter2012designing}
J.~B. Lassiter, H.~Sobhani, M.~W. Knight, W.~S. Mielczarek, P.~Nordlander, and
  N.~J. Halas, {\protect\JournalTitle{Nano letters}} \textbf{12}, 1058 (2012).

\bibitem{dregely2011excitation}
D.~Dregely, M.~Hentschel, and H.~Giessen, {\protect\JournalTitle{ACS nano}}
  \textbf{5}, 8202 (2011).

\bibitem{liu2018resonance}
S.-D. Liu, J.-L. Fan, W.-J. Wang, J.-D. Chen, and Z.-H. Chen,
  {\protect\JournalTitle{ACS Photonics}} \textbf{5}, 1628 (2018).

\bibitem{johnson1972optical}
P.~B. Johnson and R.-W. Christy, {\protect\JournalTitle{Physical review B}}
  \textbf{6}, 4370 (1972).

\bibitem{smith1985handbook}
D.~Smith, E.~Shiles, M.~Inokuti, and E.~Palik, \emph{Handbook of optical
  constants of solids}, vol.~1 (Academic Press, Inc, 1985).

\bibitem{bachelier2008fano}
G.~Bachelier, I.~Russier-Antoine, E.~Benichou, C.~Jonin, N.~Del~Fatti,
  F.~Vall{\'e}e, and P.-F. Brevet, {\protect\JournalTitle{Physical review
  letters}} \textbf{101}, 197401 (2008).

\bibitem{chong2014observation}
K.~E. Chong, B.~Hopkins, I.~Staude, A.~E. Miroshnichenko, J.~Dominguez,
  M.~Decker, D.~N. Neshev, I.~Brener, and Y.~S. Kivshar,
  {\protect\JournalTitle{Small}} \textbf{10}, 1985 (2014).

\bibitem{novotny2006principles}
L.~Novotny and B.~Hecht, \emph{Principles ofnano-optics} (Cambridge University
  Press, 2006).

\end{thebibliography}

\bibliographyfullrefs{ref}

\end{document}